\documentstyle[12pt,a4]{article}
\normalsize
\def\simless{\mathbin{\lower 3pt\hbox{$\rlap{\raise 5pt\hbox{$\char'074$}}
\mathchar"7218$}}}
\def\simgreat{\mathbin{\lower 3pt\hbox{$\rlap{\raise 5pt \hbox{$\char'076$}}
\mathchar"7218$}}}
\catcode`@=11

\def\beqra{\begin{eqnarray}} \def\eeqra{\end{eqnarray}}
\def\beq{\begin{equation}}      \def\eeq{\end{equation}}

\def\fo{\hbox{{1}\kern-.25em\hbox{l}}}

\def\ch{\@startsection{section}{1}{\z@}{-3ex plus-1ex minus-.2ex}%
        {2ex plus.2ex}{\large\sc}}

\def\; \lapp \;{\raisebox{-.4ex}{\rlap{$\sim$}} \raisebox{.4ex}{$<$}}
\def\gapp{\raisebox{-.4ex}{\rlap{$\sim$}} \raisebox{.4ex}{$>$}}
\def\con{\ifmmode \hbox{\bf*} \else{\bf*}\fi}   
\def\scon{\ifmmode \hbox{\footnotesize\rm\bf*} \else{\footnotesize\rm\bf*}\fi}

\def\0#1{\relax\ifmmode\mathaccent"7017{#1}
        \else\accent23#1\relax\fi}              

\def\eslash{\not{\hbox{\kern-2pt $E$}}}

\catcode`@=12

\begin{document}
\hoffset=0.4cm
\voffset=-1truecm
\normalsize
\pagestyle{empty}
\def\ni{{\bar {N_i}}}    \def\nj{{\bar {N_j}}}   \def\n3{{\bar {N_3}}}
\def\li{\lambda_i}    \def\lj{\lambda_j}   \def\l3{\lambda_3}
\def\hn{h^\nu}       \def\hnij{h^\nu_{ij}}
\baselineskip=5pt
\begin{flushright}
CERN-TH 6656/92
\end{flushright}
\begin{flushright}
DFPD 92/TH/43
\end{flushright}
\begin{flushright}
SISSA-127/92/AP
\end{flushright}
\vspace{24pt}
\begin{center}

{\Large {\bf The Supersymmetric Singlet Majoron}}

\vspace{24pt}

G.F. Giudice$^{1}$, A. Masiero$^{2,\,\clubsuit}$,
M. Pietroni$^{1,3}$
and A. Riotto$^{4}$
\end{center}
\vskip 0.5 cm
\centerline{\it $^{1}$ Istituto Nazionale di Fisica Nucleare,}
\centerline{\it Sezione di Padova, 35100 Padova, Italy.}
\vskip 0.2 cm
\centerline{\it $^{2}$ Theory Division, CERN,}
\centerline{\it 1211 Geneva 23, Switzerland.}
\vskip 0.2 cm
\centerline{\it $^{3}$ Dipartimento di Fisica Universit\`a di Padova,}
\centerline{\it Via Marzolo 8, 35100 Padova, Italy.}
\vskip 0.2 cm
\centerline{\it $^{4}$ International School for Advanced Studies, SISSA,}
\centerline{\it via Beirut 2-4, I-34014 Trieste, Italy.}
\vspace{36pt}
\centerline{\large{\bf Abstract}}
\vskip 0.15 cm
\baselineskip=24pt

We study the supersymmetrized version of the singlet majoron model and,
performing an analysis of the renormalization group equation improved
potential, we find that a spontaneous breaking of $R$-parity can be achieved
for a wide range of the parameters. Studying the finite temperature
effective potential, we show that the phase transition leading to $R$-parity
breaking can be of the first order and can occur at temperatures below the weak
scale, thus avoiding any constraint coming from the requirement of the
preservation of the baryon asymmetry in the early Universe.\\
--------------------------------------- \\
$^{\clubsuit}$On leave of absence from INFN, Padova, Italy.\\
CERN TH\\
September 1992
\newpage

\hoffset=0.4cm
\voffset=-1truecm
\normalsize
\baselineskip=24pt
\setcounter{page}{1}
\def\ni{{\bar {N_i}}}    \def\nj{{\bar {N_j}}}   \def\n3{{\bar {N_3}}}
\def\li{\lambda_i}    \def\lj{\lambda_j}   \def\l3{\lambda_3}
\def\hn{h^\nu}       \def\hnij{h^\nu_{ij}}
\voffset=-1 truecm
\normalsize
\begin{center}
{{\Large {\bf 1. Introduction}}}
\end{center}
\vspace{1. truecm}

The possibility that lepton number ($L$) is not an exact symmetry is of
great interest, since it leads to the prediction of non-vanishing
neutrino masses with important implications for particle
physics, astrophysics, and cosmology. Theoretically one can
envisage two schemes for $L$ violation: either $L$ is an approximate
symmetry of the standard model explicitly broken, as in
Grand Unified Theories (GUT's)-inspired see-saw mechanisms \cite{seesaw}, or
it is an exact
global symmetry spontaneously broken, leading to the existence
of a Goldstone boson, the Majoron \cite{maj}--\cite{majsing}.

In supersymmetry, non-conservation of $L$ presents novel and
characteristic features. First of all, because
of the connection between $L$ and
$R$-parity, the discrete symmetry that prevents the lightest
supersymmetric particle from decaying, the breaking of $L$ has
also important consequences for supersymmetric phenomenology.
Furthermore, unlike
the standard model, $L$ non-conserving
renormalizable interactions are allowed by gauge invariance
in supersymmetric models,
even with a minimal choice of fields,
providing a new source of explicit $L$ violation \cite{rbreak}.
Spontaneous $L$-breaking is also possible in supersymmetry
without introducing additional fields, since the scalar
partner of the neutrino may acquire a non-vanishing vacuum
expectation value \cite{rspont}. Then the Majoron is mainly the
supersymmetric partner of the neutrino and should be detected in
$Z^{0}$ decays. Recent LEP measurements \cite{lep} of the $Z^{0}$ width
have ruled out this possibility.
Nevertheless, if the spontaneous $L$-breaking is triggered by the
vacuum expectation value of the gauge-singlet right-handed
neutrino, the existence of the Majoron is not in contradiction
with the $Z^{0}$ width measurements. The feasibility of such a scheme
was first shown in the model of ref. \cite{valle} by introducing
seven new gauge
singlet superfields, besides the usual particle content of the minimal
supersymmetric standard model (MSSM).
We extend the MSSM by simply including right-handed neutrino chiral
superfields ($\hat{N}_{i}$) and an additional gauge singlet superfield
$\hat{\Phi}$, carrying two units of $L$. The most general superpotential
invariant under gauge symmetry and $L$ is
\beqra
f&=&h^d_{ij}\hat{Q}_i{\hat{d}^{c}_j}\hat{H}_1
+ h^u_{ij}\hat{Q}_i{\hat{u}^{c}_j}\hat{H}_2
+ h^e_{ij}\hat{L}_i{\hat{e}^{c}_j}\hat{H}_1
+ \hnij \hat{L}_i{\hat{N}_j}\hat{H}_2\nonumber \\
&-&\mu \hat{H}_1\hat{H}_2+\lambda_{ij}\hat{N}_{i}\hat{N}_{j}\hat{\Phi} ,
\eeqra
where $i$, $j$ are the generation indices, $\hat{L}$, $\hat{Q}$ are the
left-handed lepton and quark doublets respectively; $\hat{e}^{c}$,
$\hat{u}^{c}$ and $\hat{d}^{c}$ are the (charge conjugate of)
left-handed lepton and charge $-2/3$ and $1/3$ quark singlet superfields,
respectively; $\hat{H}_{1}$ and $\hat{H}_{2}$ are the two Higgs doublets
necessary to give masses to leptons and quarks through the Yukawa couplings
$h_{ij}^{e}$, $h_{ij}^{u}$ and $h_{ij}^{d}$.

The superpotential $f$ contains
the usual terms of the MSSM augmented by Yukawa interactions for
the right-handed neutrinos (with couplings $\hnij$) and an
interaction term for the new gauge singlet $\hat{\Phi}$ (with couplings
$\lambda_{ij}$). Our purpose in the present paper is to show
that the supersymmetric version of the singlet Majoron model
\cite{majsing} given in eq. (1) is a viable, and most economical, model for
spontaneously broken $L$-symmetry and $R$-parity. In particular, we show that
for a wide range of parameters a radiative breaking of $R$-parity can be
obtained, as derived from the analysis of the renormalization group equations
(RGE's), which we solve in two limiting cases. After a brief discussion of the
phenomenological consequences of the $L$-number spontaneous symmetry breaking,
giving rise to constraints on the $\hnij$ couplings coming from Majoron
interactions and neutrino masses, we include the one-loop finite temperature
corrections to the effective potential. This allows us to discuss
the problem of baryogenesis in this class of models with spontaneous
$R$-breaking. It is well known that instanton effects
in the Standard Model violate $B$ and $L$ while conserving the combination
$(B-L)$. There is a widespread consensus that they become
important at temperatures above $M_{W}$ \cite{sphal}. If, in addition to these
$B$- and $L$-violating effects, other interactions exist which violate $B$,
$L$, and also the combination $(B-L)$, and they are in equilibrium at
temperatures above $M_{W}$,
then no cosmological baryon asymmetry $\Delta B$ can survive \cite{campbell}.
If $R$-parity is broken explicitly \cite{rbreak}, the $L$-violating
interactions may wash out any preexisting baryon asymmetry, unless either
very strong bounds are imposed on the $R$-breaking couplings \cite{campbell},
or new mechanisms to generate or preserve $\Delta B$ are invoked, see refs.
\cite{hall,riotto} and \cite{dreiner}.
Studying the phase transition leading to the spontaneous breaking of $L$,
we show that it may take place after the completion of the electroweak
phase transition, when sphaleron interactions are frozen out and no
wiping out of a net baryon asymmetry can occur any longer.

The paper is organized as follows. In Sect. 2 we present the model and
discuss the minimization of the RGE's improved scalar potential. In Sect. 3
a brief discussion of the phenomenology of the model is presented. We proceed
to the analysis of the one-loop finite temperature corrected potential in
Sect. 4. Finally, in Sect. 5 we discuss the results and present our
conclusions.
\vspace{1.5 truecm}\\
\begin{center}
{\Large {\bf 2. Minimization of the effective potential.}}
\end{center}
\vspace{1. truecm}

The tree level potential for the neutral scalar fields $H_1^0$, $H_2^0$,
$\nu_i$ (respectively belonging to the weak doublet superfields $\hat{H}_1$,
$\hat{H}_2$,
$\hat{L}_i$) and $\Phi$, $N_{i}$ (belonging to the gauge singlet superfields)
can be decomposed as
\beq
V^{tree}=V_H^{tree}(H_1^0,H_2^0)+V_{N \Phi}^{tree}(N_{i} ,\Phi )+
V_\nu^{tree} (\nu_i , N_{i} ,\Phi ,H_1^0,H_2^0) .
\eeq
$V_{H}^{tree}$ is the usual MSSM Higgs potential. Assuming for simplicity all
$\lambda_{ij}$ real, we can work in a basis for the $N_i$ fields
where the couplings $\lambda_{ij}$ are diagonal ($\lambda_{ij}=
\li \delta_{ij}$) and write
\beqra
V_{N \Phi}^{tree}(N_i ,\Phi )&=&4\sum_i \left| \li N_i \Phi \right|^2+
\left| \sum_i \li N_{i}^{2} \right|^2 + \sum_i m_{N_i}^2 \left|
N_i \right|^2 + m_{\Phi}^2 \left| \Phi \right|^2 \nonumber \\
&-& \left( \sum_i
A_i \li N_{i}^{2} \Phi +\rm{h.c.}\right) ,
\eeqra
where the last three terms contain the soft supersymmetry breaking
masses ($m_{N_i} ,m_\Phi$) and trilinear couplings ($A_i$). \\
It is
plausible to expect that the coupling constants $\hn$ are of the
same order of magnitude of $h^e$, the Yukawa couplings for the
charged leptons, and thus numerically small (\mbox{$h^e
\simeq 10^{-2}-10^{-6}$}). Assuming then that $|\hn|\ll 1$ and keeping only the
leading terms of an expansion in $\hn $, we obtain
\beqra
&&V_{\nu}^{tree} (\nu_i , N_i ,\Phi ,H_{1}^{0},H_{2}^{0})=
\frac{g^2+g'^2}{8}\left[\left( \sum_i \left|\nu_i \right|^2 \right)^2\right.
\nonumber
\\ && \left. +2 \sum_i \left|\nu_i \right|^2\left(\left|H_{1}^{0}\right|^2 -
\left|H_{2}^{0}\right|^2 \right)\right] +
\sum_i m_{\nu_i}^2 \left| \nu_i \right|^2\\ &&+\left[ \sum_{ij}
\hnij \nu_i \left( 2\lj N_{j}^{\ast} \Phi^\ast H_2- \mu N_j
{H_{1}^{0}}^{\ast} -A^{(h)}_{ij} N_j H_{2}^{0}\right) + \rm{h.c.}\right]
+ {\cal O}({\hn}^2 ) \nonumber ,
\eeqra
where we have again introduced the appropriate soft supersymmetry
breaking mass parameters ($m_{\nu_i}$) and trilinear couplings
($A^{(h)}_{ij}$), and $g$ and $g'$ are the $SU(2)_L$ and $U(1)_Y$ gauge
couplings, respectively.

As is usually done for the MSSM, we will assume that the supersymmetry
breaking terms have a common origin and are therefore related
at some GUT scale ($M_{GUT} \simeq 10^{16}$ GeV), such as all mass
parameters and trilinear couplings are equal to a universal mass
$\tilde {m}$ and a universal coupling $A$. The values of the
supersymmetry breaking parameters appearing in the scalar potential,
\mbox{eqs.(2-4)}, are then derived by solving the relevant RGE's with boundary
conditions at $M_{GUT}$.
It is well known that, for an appropriate choice of initial
conditions at $M_{GUT}$, the effect of running the parameters
in $V_{H}^{tree}(H_1^0,H_2^0)$ from $M_{GUT}$ to low energy
is to drive electroweak symmetry breaking
with non-vanishing vacuum expectation values for the Higgs fields,
$\langle H_1^0 \rangle =\frac{v}{\sqrt{2}} \cos \beta$, $\langle H_2^0 \rangle
=
\frac{v}{\sqrt{2}} \sin \beta$,
with $v=246$ GeV.

We will minimize the renormalization group improved potential (3) and safely
neglect the logarithmic term ${\rm Str}{\cal M}^4{\rm ln} ({\cal M}^2/Q^2 )$ by
choosing the renormalization scale $Q$ at the order of the weak scale.
Moreover, assuming ${\cal CP}$-conservation, we may take all the parameters
in eq. (3) to be real.

By assuming that the $\langle \nu_i \rangle$' s are much smaller
than the typical weak scale $v$, we can neglect the quartic terms in eq. (4).
We obtain the approximate solution\footnote{Here, in order to avoid
non-vanishing $\langle \nu_i \rangle$'s for $\hn =0$, we are assuming
$m_{\nu_i}^2+(M_Z^2/2)\cos 2\beta > 0$. This is however not very
restrictive, since it holds for almost all of the supersymmetric
parameters which lead to a correct electroweak symmetry breaking.}:
\beq
v_{\nu_i} = v \frac{\sum_j \hnij v_{N_j} \left[ \mu \cos \beta +
(A^{(h)}_{ij} -2\lj v_\Phi )\sin \beta \right] }{\sqrt{2} \left[m_{\nu_i}^2 +
(M_Z^2/2)\cos 2\beta\right]} + {\cal O}({\hn}^2),
\eeq
where $v_\Phi \equiv \langle \Phi \rangle$, $v_{N_i} \equiv \langle
N_i\rangle$.
Notice that, for consistency, we must require  $v_{\Phi}\simeq v_{N_{i}}
={\cal O}(1)$ TeV and $\hn \ll 1$.

The values of $v_\Phi$ and $v_{N_{i}}$ are determined by the minimum of
$V_{N\Phi}^{tree}$, eq. (3)
\beq
v_\Phi=\frac{x}{4\lambda_{3}},~~~~~~v_{N_{3}}^2=\frac{m_\Phi^2}{4\lambda_{3}^2}
\frac{x}{(A_{3}-x)},~~~~~~v_{N_2}=v_{N_1}=0,
\eeq
where $x$ is a solution of the cubic equation:
\beq
x^3-3A_{3}x^2+2\left( 2m_{N_{3}}^2-m_\Phi^2+A_{3}^2\right) x-4A_{3}
m_{N_{3}}^2=0,
\eeq
under the condition
\beq
A_{3}/x>1.
\eeq
We have chosen $\lambda_{3}$ to be the smallest of the non-zero values of
$\li$.
In the case $\lambda_1=\lambda_2=\l3$, a minimum with two non-vanishing
$v_{N_i}$ is possible.
However, such a minimum requires $m_\Phi^2>m_{N_i}^2$, in contradiction
with the solution of the RGE's, see eqs. (21-22) and the Appendix.

The potential $V_{N \Phi}^{tree}$ at the minimum is
\beq
V_{N \Phi}^{tree}(v_{N_i} ,v_\Phi )=\left( \frac{m_\Phi x}{4\lambda_{3}}
\right)^2
\left[ 1-\left( \frac{m_\Phi}{A_{3}-x}\right)^2 \right] .
\eeq
The condition for spontaneous $L$-breaking is that
$V_{N \Phi}^{tree}(v_{N_i} ,v_\Phi )$
is smaller than the value of the potential at the origin:
\beq
A_{3}-m_\Phi <x<A_{3}+m_\Phi .
\eeq
Moreover the requirement that the potential $V_{N \Phi}^{tree}$ is bounded
from below implies the condition:
\beq
m_\Phi^2>0.
\eeq

As stated above, the parameters in the potential $V_{N \Phi}^{tree}$, eq. (3),
are related by an initial condition at $M_{GUT}$. Their dependence
upon the energy scale $Q$ is determined by the following
RGE's ($t\equiv \log Q/M_{GUT}$)
\beqra
\frac{d}{dt}\li&=&\frac{5}{8\pi^2}\li^3, \\
\frac{d}{dt}m_{N_i}^2&=&\frac{1}{2\pi^2}\left( 2m_{N_i}^2+m_\Phi^2+|A_i|^2
\right) \li^2, \\
\frac{d}{dt}m_\Phi^2&=&\frac{1}{4\pi^2}\sum_i
\left( 2m_{N_i}^2+m_\Phi^2+|A_i|^2 \right) \li^2, \\
\frac{d}{dt}A_i&=&\frac{1}{4\pi^2}\left( 4A_i\li^2+\sum_jA_j\lj^2
\right),
\eeqra
with boundary conditions at $Q=M_{GUT}$: $m_{N_i}^2(0)= m_\Phi^2(0)=
{\tilde m}^2$, $A_i(0)=A$, $\li(0)=\li^{(0)}$.

Eqs. (12-15) can be easily solved either in the limit of vanishing
$\lambda_1$ and $\lambda_2$:
\begin{equation}
\begin{array}{lccclcc}
\lambda_{1,2}(Q)&=&0,&&\l3 (Q)&=&\l3^{(0)}K,\\
A_{1,2}(Q)&=&AK^{2/5},&&A_3(Q)&=&AK^2,
\end{array}
\end{equation}
\beqra
m_\Phi^2(Q)&=&\frac{\tilde m^2}{5}\left[ 2+(3-A^2)K^2+A^2K^4\right] ,\\
m^2_{N_{1,2}}(Q)&=&{\tilde m^2},~~~~~~~~~~~~~
m_{N_3}^2(Q)=2m_\Phi^2(Q)-{\tilde m^2},
\eeqra
\beq
K\equiv \left( 1+\frac{5}{4\pi^2} \l3^{(0)2}\log
\frac{M_{GUT}}{Q}\right)^{-1/2}=
        \left( 1-\frac{5}{4\pi^2} \l3^2(Q)\log
\frac{M_{GUT}}{Q}\right)^{1/2},
\eeq
or in the limit $\lambda_1=\lambda_2=\l3$:
\beqra
\li (Q)&=&\li^{(0)}K,~~~~~~~~~A_i (Q)=AK^{14/5},\\
m_\Phi^2(Q)&=&\frac{\tilde m^2}{7} \left[ -2+3(3-A^2)K^{14/5}
+3A^2K^{28/5}\right],\\
m_{N_i}^2(Q)&=&\frac{2}{3}m_\Phi^2(Q) +\frac{1}{3}{\tilde m^2}.
\eeqra

In summary,
$L$-symmetry and $R$-parity are spontaneously broken if eq. (7) allows
a solution which satisfies the conditions (8), (10), and (11), with
the supersymmetry breaking parameters fixed at low energy by the
RGE's (12-15). In figs. 1-2 we show the region of parameters
$\l3 (Q^2=M_W^2)$ and $A_3 (Q^2=M_W^2)$ where $L$ is spontaneously
broken in the case {\it a)} $\lambda_1=\lambda_2=0$, $\l3 \neq 0$
(where eqs. (16-19) can be applied\footnote{In such a case only one
gauge singlet superfield $\hat{N}$ is responsible for all the light
neutrino masses.})
and {\it b)} $\lambda_1=\lambda_2
=\l3$ (where eqs. (20-22) can be applied). An upper bound on $\left| A\right|$
follows from the constraint of avoiding undesirable color-breaking
minima \cite{color}
\beq
|A|^2<3(3{\tilde m^2}+\mu^2 ),
\eeq
at the GUT scale. Since the value of $\mu$, the Higgs mixing parameter,
see eq. (1), does not explicitly enter in
our analysis, we have chosen to plot in
figs. 1-2 the constraint $|A|<5{\tilde m}$, as representative of eq. (23).
For small values of
$\li$, eq. (7) can be solved analytically and the solution which
satisfies conditions (8), (10), and (11), at the weak scale, is
\beq
x=\frac{A_{3}}{2}\left( 1+\sqrt{1-\frac{8 \tilde{m}^2}{A_{3}^2}}\right)
{}~~~~~~{\rm
for}~~~~|A_{3}|>3 {\tilde m}.
\eeq
In the case {\it a)}, for larger values of $\l3$ ($\l3 \gapp 0.1$),
solutions for smaller values of $\left |A_3\right|$ are possible, and they
correspond mainly to parameters such that $m_{N_{3}}^2<0$, see fig. 1.
In the case
{\it b)} such solutions are not present, since the RGE's
imply now $m_\Phi^2<m_{N_{i}}^2$ and the condition (11) cannot be
satisfied if $m_{N_{i}}^2<0$. In the case {\it b)} there is also an
approximate (broken by Yukawa terms) $O(3)$ generation symmetry of
the right-handed neutrinos, which is spontaneously broken to
$O(2)$, giving rise to two pseudo-Goldstone bosons (see Appendix). This is not
necessarily a problem, since these light familons would be very
weakly coupled to ordinary particles.
\vspace{1.truecm}\\
\begin{center}
{\Large{\bf 3. Phenomenology of the model.}}
\end{center}
\vspace{1. truecm}

The spontaneous breaking of $L$-symmetry will give rise to a
Goldstone boson, the Majoron. Since the $\langle \nu_i \rangle$'s carry
both $L$ and hypercharge quantum numbers, the physical Goldstone
boson is given by:
\beqra
J&\simeq& {\rm Im} \left[ \left( \frac{\sum_i v_{\nu_i}^2}{vv_L}\right)
\sqrt{2}(\sin \beta H_2^0 -\cos \beta H_1^0)\right.\nonumber\\
&+&\left.\sum_i \left( \frac{v_{\nu_i}}
{v_L}\nu_i-\frac{v_{N_{i}}}{v_L}N_{i} \right) +2\frac{v_\Phi}{v_L} \Phi
\right],
\eeqra
\beq
v_L^2\equiv 4v_\Phi^2+\sum_i v_{N_{i}}^2,
\eeq
where we have used the fact that the $v_{\nu_i}$'s are much smaller
than the other vacuum expectation values.

The strongest bounds on the Majoron couplings come from the observed
duration of helium burning in red-giants. Ref. \cite{raff} gives
\beq
g_e \simless 3 \times 10^{-13} ~~~~~~{\rm and}~~~~~~g_\gamma\simless 10^{-10}
{}~ {\rm GeV}^{-1},
\eeq
where $g_e$ is the coupling constant of the pseudoscalar interaction
of the Majoron with electrons and $g_\gamma$ is the coupling constant
of the Majoron-photon interaction term
\beq
-\frac{g_\gamma}{4}F^{\mu \nu}
{\tilde F}_{\mu \nu} J.
\eeq

Since the $L$ current is anomalous, a coupling of the form (28) is
generated at the one loop level by a triangular graph with charginos
and charged leptons running inside the loop. Although the exact
result for $g_\gamma$ depends on the particular choice of the
supersymmetric parameters, we can give an estimate of it. We first
assume that the Yukawa couplings $\hnij$ have a hierarchical
structure in generation indices (like $h^e_{ij}$), with
$\hn_{31}\simless\hn_{32}\simless\hn_{33}< 1$.
Taking now the supersymmetric parameters
of the same order of magnitude as the weak scale and all
coupling constants, except Yukawas, of order unity, the one-loop induced
Majoron-photon coupling is:
\beq
g_\gamma \simeq \frac{\alpha}{2\pi}\frac{{\hnij}^2}{v_L},
\eeq
and from eq. (27) we obtain the bound $\hn_{33} \, \simless \,  10^{-2}
\sqrt{v_L/{\rm TeV}}$.

A more stringent bound comes from the Majoron-electron coupling.
Under the same assumptions as before, we can estimate $g_e\simeq
{\hn_{33}}^2h^e_{11}$ from the Higgs component of the Majoron, and
$g_e \simeq {\hn_{13}}^2$ from the scalar neutrino component of
the Majoron. Barring the possibility of accidental cancellations,
eq. (27) implies $\hn_{33}\, \simless \, 0.4\times 10^{-3}$ and
$\hn_{13}\, \simless \, 0.5\times 10^{-6}$. Of course this is to be understood
as merely an estimate of the bound, which actually depends on the
various input parameters of the MSSM. However it is interesting to
observe that $\hn_{33}\simeq 10^{-3}$ and $\hn_{13}\simeq 10^{-6}$
are the values suggested by an analogy with the charged lepton
Yukawa couplings.

After spontaneous symmetry breaking, neutrinos acquire masses. The
three families of left-handed neutrinos mix with the right-handed
neutrinos and with the neutralinos in a $11\times 11$ mass matrix.
The neutrino masses are therefore complicated functions of the
supersymmetric parameters and coupling constants. However, in the
above-mentioned approximation of keeping a single energy scale
for the supersymmetric parameters and keeping all the coupling constants,
except
the Yukawas, equal to one, we find
\beq
m_{\nu_\tau}\simeq {\hn_{33}}^2 \tilde{m}.
\eeq
For $\hn_{33}
\simeq 10^{-3}$ and $\tilde{m}\simeq$ 1 TeV, the tau neutrino mass is
$m_{\nu_\tau}\simeq$ 1 MeV, an interesting
value for phenomenology. The other two
neutrinos $\nu_e$ and $\nu_\mu$ are expected to be almost massless,
since the mass matrix has two approximate zero eigenvalues (even for
$\hn_{13},~ \hn_{23} \neq 0$). The massive $\nu_\tau$ does not
cause cosmological problems since it is short lived
\beq
\tau (\nu_\tau \to \nu_i J)\simeq \frac{16 \pi}{m_{\nu_\tau}}
{\hn_{33}}^2 {\hn_{i3}}^2\simeq 3\times 10^{-6}
\left( \frac{\hn_{33}}{10^{-3}}\right)^{-4}
\left( \frac{\hn_{i3}}{10^{-4}}\right)^{-2}
\left( \frac{\tilde{m}}{{\rm TeV}} \right)^{-1}{\rm s}.
\eeq
Notice that the Majoron interacts with ordinary particles with
strength always proportional to $\hn$ and therefore decouples
in the limit $m_\nu \to 0$.

Even if the Majoron is coupled with ordinary matter weakly
enough to escape bounds from stellar cooling ($\hn_{33}
\, \simless \, 10^{-3},\,\hn_{13}\, \simless \, 10^{-6}$), it still could be
detected in rare $L$-violating tau and muon decays. Processes
like $\tau \to \mu J$, $\tau \to eJ$, and $\mu \to e J$ occur
with rates observable at future dedicated experiments, as has
been exhaustively discussed in ref. \cite{valtau}. On the other hand,
for $\hn \, \simless \, 10^{-3}$, the Majoron cannot be detected in
LEP experiments looking for direct Majoron production in $Z^{0}$ decays
accompanied by either $\gamma$, ${Z^{0}}^\ast$, or a light scalar boson.

Together with $L$-symmetry, $R$-parity is also spontaneously broken.
As a consequence, the lightest supersymmetric particle can decay,
modifying the standard experimental signals of supersymmetry.
Assuming that the neutralino ($\chi^0$) is the lightest supersymmetric
particle, \mbox{$\chi^0\to J \nu_\tau$}, \mbox{$\chi^0\to Z^0 \nu_\tau$},
\mbox{$\chi^0\to W^\pm \tau^\mp$} are possible decay modes. If $\chi^0$ is
heavier than $Z^0$ and $W^\pm$ gauge bosons, the three decay modes
are expected to occur at similar rates (with details depending
upon the choice of supersymmetric parameters), and the $\chi^0$
lifetime is:
\beq
\tau (\chi^0)\simeq \frac{16 \pi}{{\hn_{33}}^2 m_{\chi^0}}\simeq
\left( \frac{10^{-3}}{\hn_{33}}\right)^2 \left( \frac{300~ {\rm GeV}}
{m_{\chi^0}}\right) 10^{-19} {\rm s}.
\eeq
The two-body decay is fast, so that a $\chi^0$ produced in collider
experiments decays within the detector, unless $\hn_{33}$ is
extremely small. If $\chi^0$ is lighter and the gauge bosons
produced in the decay are virtual, $\chi^0\to J \nu_\tau$ becomes
the dominant mode, leaving no visible trace of the neutralino
decay. On the other hand, the chargino could instead be the
lightest supersymmetric particle, with a decay mode $\chi^\pm
\to J \tau^\pm$.
\vspace{1. truecm}\\
\begin{center}{\Large{\bf 4. Baryogenesis and spontaneous  $R$-breaking.}}
\end{center}
\vspace{1. truecm}

As pointed out in the Introduction, one of the major reasons for interest
in breaking $R$-parity spontaneously instead of explicitly
is the possibility of avoiding the tough constraint imposed by  the survival
of the cosmic baryon asymmetry $\Delta B$ \cite{campbell}.
To make this point clear, let us
first review why baryogenesis so severely constrains {\it explicit}
$R$-breaking.

The absence of $R$-symmetry in the initial superpotential
implies that either $L$- or $B$- number is explicitly violated (in principle,
both these two numbers could be violated, but, in that case, an unbearably
fast proton decay would occur). If the $L$- or $B$-violating interactions,
which derive from these $R$-violating terms in the superpotential, are in
thermal equilibrium at temperatures for which the $(B+L)$-violating anomalous
electroweak interactions are operative, any possible preexisting $\Delta B$
will be erased. Requiring the $R$-violating
interactions to be constantly out of equilibrium implies very stringent
upper bounds on their couplings in the superpotential of the order of
$10^{-7}-10^{-8}$. Clearly, if this is the case,
$R$-violating effects are completely invisible in accelerator physics
experiments.

Two possible loopholes have been proposed in the literature:
$i)$ $R$-breaking in the leptonic
 sector leaves some partial lepton number unbroken ($L_{unb}$), so that
$(\frac{1}{3} B-L_{unb})$ replaces the usual $(B-L)$ number \cite{dreiner};
$ii)$ some new
mechanism for a late $\Delta B$ production at the Fermi scale is operative.
In this latter case, one can try to make use of an interesting interplay
between $R$-violating interactions in the leptonic sector, giving rise
to a net $\Delta L$ at the weak scale, and the still operative
anomalous interactions,
converting this $\Delta L$ into a net $\Delta B$ \cite{riotto}.

If $R$-parity is an exact symmetry to start with, and only subsequently broken
in a spontaneous way, then we can envisage what we consider the most attractive
resolution of the conflict between $R$-breaking and baryogenesis. Namely,
$R$ could remain a good symmetry throughout the whole interval between the
Planck and Fermi scales, exhibiting a spontaneous breaking at temperatures so
low that sphalerons are no longer operative. If $R$ is broken in such a way
in the leptonic sector, there is obviously no harm at all for any preexisting
$\Delta B$. The crucial question to answer is then: is it possible to
achieve a spontaneous breaking of $L$ at a temperature below that at
which the electroweak phase transition occurs? We will show that this is
possible at least for a range of the $\lambda$ parameters of our model.

First we include the one-loop finite temperature corrections into the
potential of eq. (3) \cite{dolan}. In both cases {\it a)} and {\it b)},
see Sect. 2,
only one out of the three fields $N_i$, henceforth $N$, acquires a VEV,
so that we reduce our analysis along the $N$ and $\Phi$ directions.

In the limit $\hn\rightarrow 0$ and in terms of $N_{R}={\rm{Re}}\,N/\sqrt{2}$,
$\phi_{R}={\rm{Re}}\,\Phi/\sqrt{2}$, the finite temperature contribution to
the effective potential is
\begin{eqnarray}
V_{N \Phi}^{T}\left(N_{R}, \phi_{R}\right) & = & \frac{1}{2} \mu_{N}^{2}(T)
N_{R}^{2} +
\frac{1}{2} \mu_{\Phi}^{2}(T) \phi_{R}^{2} - \frac{T}{12\pi}\sum_{i\,B}
m_{i}^{3} \nonumber \\
 & + & \sum_{i\,F} N_{i}\frac{m_{i}^{4}\left(N_{R},\phi_{R}\right)}{64 \pi^{2}}
{\rm ln} \left[ \frac{m_{i}^{2} \left(N_{R},\phi_{R}\right)}{A_{f}
T^{2}}\right]
\nonumber \\
& - & \sum_{i\,B} N_{i}\frac{m_{i}^{4}\left(N_{R},\phi_{R}\right)}{64 \pi^{2}}
{\rm ln} \left[ \frac{m_{i}^{2} \left(N_{R},\phi_{R}\right)}{A_{b} T^{2}}
\right],
\end{eqnarray}
where
\begin{eqnarray}
\mu_{N}^{2}(T) & = & m_{N}^{2} + \lambda^{2} T^{2}, \nonumber\\
\mu_{\Phi}^{2}(T) & = & m_{\Phi}^{2} + \frac{1}{2}\lambda^{2} T^{2},
\end{eqnarray}
are the $N$ and $\Phi$ plasma masses; $m_{i}$'s are the mass eigenvalues of
any particle
with $N_{i}$ degrees of freedom, $A_{f}= \pi^{2} {\rm exp}
(3/2-2\gamma_{E})$, $A_{b}= 16 A_{f}$, and $\gamma_{E}\simeq 0.57$.\\
Note that we are assuming ${\cal CP}$-conservation so that
${\rm Im}\langle \Phi
\rangle={\rm Im}\langle N\rangle=0$ at all the temperatures.

The above expression is valid for all the temperatures $T$ larger than any
$m_{i}$,
provided that $N_{R}$ and $\phi_{R}$ are in thermal equilibrium. The
interactions coming from the terms
$\hn {\hat L}{\hat H}{\hat N}$ and $\lambda {\hat N}{\hat N} {\hat \Phi}$ in
eq. (1) keep these fields in equilibrium as long
as their rate $\Gamma$ is larger than the expansion rate of the Universe
\mbox{$H\simeq g_{*}(T) T^{2}/M_{P}$}, where $g_{*}(T)$ is the number of the
relativistic degrees of freedom present at temperature $T$ and $M_{P}$
is the Planck mass. This requirement is fulfilled for $T\simless h^{2}M_{P}$
and $T\simless \lambda^{2}M_{P}$, which, for reasonable values of
$h$ and $\lambda$, are certainly satisfied for all the temperature range of
interest.

At very high temperature, $V_{N\Phi}^{T}\left(N_{R}, \phi_{R}\right)$ is
approximated
by
\begin{equation}
V_{N\Phi}^{T}\left(N_{R}, \phi_{R}\right)\simeq
\frac{1}{2}\mu_{N}^{2}(T)N_{R}^{2}+
\frac{1}{2}\mu_{\Phi}^{2}(T)\phi^{2}_{R},
\end{equation}
so that the minimum is at $\langle N_{R}\rangle=\langle \phi_{R}\rangle=0$,
and $L$ is preserved.

Let us now consider the range $\lambda={\cal O}(1)$. As we have previously
seen,
in the case {\it a)} of Sect. 2, we found numerical
solutions for $\lambda_{3} \simgreat 0.1$ with $m_{N}^{2}$ driven to negative
values at the weak scale, as one can immediately deduce from the RGE's,
see eqs. (16-19). In such a case,
the critical temperature $T_{C}$ may be defined as the temperature at which
the effective potential develops a flat direction at the origin. From eq. (34)
one can infer that
\begin{equation}
T_{C}^{2}\simeq -\frac{m_{N}^{2}}{\lambda^{2}},
\end{equation}
and we conclude that in this case the phase
transition related to $L$-breaking can occur after the electroweak
phase transition, since $m_{N}^{2}$ can
be driven to sufficiently small values at the weak scale. Indeed, the
\mbox{$L$-violation} can occur at temperatures
low enough for the sphaleron transition to be completely out of equilibrium.
Moreover, if $T_{C}$ is small enough, the minimum of the effective potential
for $T\simeq T_{C}$ may be already formed near the vacuum at $T=0$. In fact,
if $T_{C}\ll m_{i}$, the finite temperature corrections to the effective
potential become negligible around the vacuum at $T=0$ due to an exponential
Boltzmann suppression \cite{fukugita}. In this situation the transition is
expected to be of
first order.
However, one might still wonder whether the reheating temperature at the
end of the phase
transition is high enough to reestablish the dangerous equilibrium coexistence
between anomalous electroweak interactions and $L$-violating phenomena.
We proceed to estimate the reheating temperature $T_{RH}$.
In the limit of
vanishing $m_{N}^{2}$, eq. (7) can be exactly solved and the energy at the
minimum is given by
\begin{equation}
V_{N\Phi}^{min}=-\frac{1}{64}\frac{A^{4} }{\lambda^{2}} f(k),
\end{equation}
where
\begin{equation}
f(k) = \frac{1}{2} -10 k -4 k^{2} +\frac{1}{2} (1+8 k)^{3/2},
\end{equation}
and $k\equiv (m_{\Phi}^{2} /A^{2} )$.
Eqs. (8) and (10) require $0\leq k \leq 1$, and therefore $f(k)$ varies
between 1 and 0.

Comparing eq. (37) with
\begin{equation}
\left| V_{N\Phi}^{min}\right| = \frac{\pi^{2}}{30} g_{*}(T) T_{RH}^{4},
\end{equation}
we find
\begin{equation}
T_{RH} \simeq 0.4\,g_{*}^{-1/4}(T) \frac{A }{\sqrt{\lambda}} f(k)^{1/4}.
\end{equation}
{}From the above expression we estimate that the reheating temperature can be
lower than the critical temperature of the electroweak phase transition in
the whole range for $\lambda$ under consideration, {\it i.e.} $\lambda
\simgreat 0.1$, and $k$ close to 1. For instance, for $A \simeq 1$ TeV,
$T_{RH} \simless 100$ GeV for $\lambda \simgreat 0.1$ and $k \simgreat 0.8$.
Hence, for $\lambda$ large enough, we can avoid
a sizeable reheating and the preexisting $\Delta B$ safely survives.

We are left with the other region in the parameter space where $\lambda\ll 1$.
To discuss this case it is useful to parametrize the potential in eq. (33)
in such a way which makes evidence of the dependence of the terms on $\lambda$.
This is readily achieved by defining the following dimensionless quantities
\begin{equation}
\tilde{\phi}_{R}\equiv \lambda\frac{\phi_{R}}{\tilde{m}},~~~~~~~
\tilde{N}_{R}\equiv \lambda\frac{N_{R}}{\tilde{m}},~~~~~~~
\tilde{T}\equiv \lambda\frac{T}{\tilde{m}},
\end{equation}
and
\begin{equation}
\tilde{m}_{N}^{2}\equiv \lambda^{2}\frac{m_{N}^{2}}{\tilde{m}^{2}},
{}~~~~~~~
\tilde{m}_{\Phi}^{2}\equiv \lambda^{2}\frac{m_{\Phi}^{2}}{\tilde{m}^{2}},
{}~~~~~~~ \tilde{m}_{i}^{2}\equiv \lambda^{2}\frac{m_{i}^{2}}{\tilde{m}
^{2}}.
\end{equation}
Now the effective potential becomes
\begin{equation}
\tilde{V}_{N\Phi}^{T}\left(\tilde{N}_{R},\tilde{\phi}_{R}\right)=\frac{\tilde{m}^{4}}
{\lambda^{4}}\left[ {\cal F}\left(\tilde{N}_{R},\tilde{\phi}_{R},\tilde{m}
_{N}^{2},\tilde{m}_{\Phi}^{2},\tilde{T}\right)-\frac{\lambda}{12\pi}\tilde{T}
\sum_{i\, B} \tilde{m}_{i}^{3} + {\cal O}\left(\lambda^{2}\right)\right],
\end{equation}
where ${\cal F}$ can be easily deduced from eq. (33).

Having $\lambda \ll 1$, we can neglect terms of higher order in $\lambda$ and
in this case the phase transition related to the spontaneous breaking of
\mbox{$L$-number} is likely to occur at a temperature higher than that
characteristic of the electroweak phase transition.
Indeed, from eq. (45), one can infer that
the dimensionless critical temperature has to be ${\cal O}(1)$, that is
the critical temperature is expected to be ${\cal O}(\tilde{m}/\lambda)$.
Hence, there exists an interval during which $L$ and $(B+L)$ are simultaneously
violated. We can envisage two strategies for a net $\Delta B$ to be present
at the end of the electroweak transition:\\

$i)$ the $L$-violating interactions which arise from the spontaneous breaking
of $R$ are never in equilibrium throughout the entire time sphalerons are still
active;

$ii)$ the presence of $L$-violation allows the production of a net
$\Delta B$ via sphaleron interactions.

Option $i)$ leads to a situation similar to that described at the
beginning of this Section for the case of explicit $R$-breaking.  Obviously,
also in this case it might happen that some partial lepton number remains
unbroken and then $B$ minus this number remains a good symmetry of the
theory.
In the case $ii)$, one can envisage the possibility of producing particles
with $L$-violating decays via bubble collisions at the $L$-breaking phase
transition. The distributions of particles originated
in this way are automatically out of thermal equilibrium. As for
\mbox{${\cal CP}$-violation}, the other
crucial ingredient to obtain a net $\Delta L$,
the situation is more promising than in the Standard Model,
since there are additional
couplings (the $\hn$'s) which can be complex and whose phases are not severely
limited by the existing phenomenological constraints on \mbox{${\cal CP}$-
violation}.
Provided that these $L$-violating decays are fast enough, we can invoke
processes induced by sphalerons to convert $\Delta L$ into $\Delta B$.
One can work out an explicit example making use of the production and
subsequent decay of the lightest "right-handed neutrino" $N$ to give rise
to a net $\Delta L$ along the lines of an analogous situation studied in
ref. \cite{riotto}.
The main difficulty for these scenarios is not so much
the creation of a sizeable $\Delta L$, but rather the
threat of washing out such $\Delta L$ by those same $L$-violating
interactions which gave rise to it. The point is that in the case
$\lambda\ll 1$ the $L$-breaking phase transition takes place at
a temperature of ${\cal O} (\tilde{m}/\lambda)$, high enough to produce
all the usual particles of the Standard Model through bubble collisions.
In particular, at this temperature
processes like $\nu\nu\rightarrow h h$  occur. It is easy to see that,
if some $\hn$ couplings are large enough to give rise to a sizeable
$\Delta L$ in the $N$ decays, then some neutrino annihilations into
Higgs pairs (or the inverse process) are fast enough to be in
equilibrium, hence possibly erasing the $\Delta L$. In some contrived
situations it is possible to avoid (at least) a large washing out of $\Delta
L$,
but it is likely that in the case $\lambda \ll 1$ it is quite difficult
to preserve the dynamical generation of $\Delta L$ which resulted from
the $L$-breaking phase transition. For such a reason we believe that the
case of $\lambda={\cal O}(1)$ is much more appealing as far as the
problem of preserving a net $\Delta B$ is concerned, since the limits
on the coupling constants are either absent or much less severe than
those present in models with explicit $R$-parity breaking, unless new
mechanisms are called into play, see refs. \cite{hall,riotto}
and \cite{dreiner}.
\newpage
\centerline{\Large\bf 5. Conclusions.}
\vspace{1. truecm}

The survival of the cosmic matter-antimatter asymmetry imposes severe
constraints on the $B$- and $L$-violating terms which are otherwise allowed
by the gauge symmetry and global $N=1$ supersymmetry in low energy
\mbox{supersymmetric} extensions of the Standard Model.
Imposing ordinary $R$-parity
seems to be the safest solution for this problem. However, in view of the large
ignorance surrounding the original theory from which low energy supersymmetry
arises, we think that it is worthy to study how alternatives with explicit or
spontaneous violations of $R$-parity address the problem of the $\Delta B$
survival. For models with Baryon Parity \cite{ibanez}, {\it i.e.} a discrete
symmetry
which preserves $B$ while allowing for $L$-violation in the superpotential,
the best way to avoid the washing out of $\Delta B$ seems to be the
conservation
of a residual lepton-flavour number \cite{dreiner}. Alternatively,
one can try to produce
a new $B$ asymmetry at the electroweak phase transition through an articulated
interplay of $L$- and $(B+L)$-violating interactions \cite{riotto}. However
this late dynamical production of $\Delta B$ depends on the bubble dynamics
which is still a rather open subject.

In this paper we have focused on the survival of $\Delta B$ in models where the
initial supergravity Lagrangian possesses $R$-symmetry, which is subsequently
broken in a spontaneous way at a temperature of the order of the Fermi energy
scale. We have related the spontaneous breaking of $R$-parity to that of the
lepton number $L$.
After LEP results on the $Z^{0}$ lineshape the only available
possibility seems to be through the VEV of a scalar which is singlet under
$SU(2)_{L}\otimes U(1)_{Y}$. Hence, the simplest extension  of the MSSM which
allow for such a breaking of $L$ includes some new singlet superfield carrying
$L$. We have studied here the possibility of supersymmetrizing the original
singlet majoron model of Chikashige, Mohapatra and Peccei \cite{majsing}.
We have showed that it is possible to achieve a radiative breaking of $R$
for a wide range of parameters.
This class of models with radiative breaking of $L$ exhibits a
crucial feature for the problem of the survival of the cosmic $\Delta B$:
we can find initial conditions allowing for a very late radiative violation of
$L$ at temperatures such that sphaleron induced processes are no longer
operative. On the other hand, some choices of these initial conditions allow
for effects of broken $R$-parity potentially accessible in accelerator
experiments.

In conclusion, the claim that an observable $R$-parity violation implies
a complete washing out of the cosmic $\Delta B$ finds an interesting way out:
it is possible to construct classes of supersymmetric theories where $R$-parity
remains a good symmetry throughout all the history of the early Universe
down to temperatures of ${\cal O}(M_{W})$, being subsequently broken in a
spontaneous way by radiative effects when the anomalous $(B+L)$-violating
effects are out of equilibrium. The concrete, phenomenologically viable
illustration that we provide here should stimulate further effort in the
direction of a cosmologically acceptable $R$-violation with visible
effects at collider physics.

After accomplishing our work, we received the paper "Phase Transition
for $R$-parity Breaking" by M. Chaichian and R. Shabad in which the same
point of a late spontaneous $R$ breaking is illustrated in a different
supersymmetric realization.

\vspace{1.truecm}
\centerline{\Large\bf Acknowledgements}
\vspace{0.5 truecm}

We thank F. Feruglio for many useful discussions and for participating in
some stages of this work. Two of us (M.P. and A.R.) would like to thank the
Laboratori Nazionali del Gran Sasso, where the present work was completed,
for the kind hospitality.
K. Enqvist is also acknowledged for useful discussions.
\newpage
\centerline{\Large\bf Appendix.}
\vspace{1. truecm}

 In this Appendix we wish to discuss in more detail the minimization
of the effective potential in the case $\lambda_{1}=\lambda_{2}=\lambda_{3}
\equiv \lambda$, such that $A_{1}=A_{2}=A_{3}\equiv A$.

Making use of cartesian coordinates
\[
\left\{\begin{array}{ccc}
N_{i}&=& (X_{i}+i Y_{i})/\sqrt{2},\\
 & & \\
\Phi &=& (F+iG)/\sqrt{2},
\end{array}\right.
\]
\begin{flushright}
{(A.1)}
\end{flushright}
the effective potential, see eq. (3), reads
\beqra
V_{N\Phi}^{tree}&=& \lambda^{2}(F^{2}+G^{2})(\vec{X}^{2}+\vec{Y}^{2})+
\frac{\lambda^{2}}{4}\left[(\vec{X}^{2}-\vec{Y}^{2})^{2} +4(\vec{X}
\cdot \vec{Y})^{2}\right]\nonumber \\
&+& \frac{m_{N}^{2}}{2}(\vec{X}^{2}+\vec{Y}^{2}) + \frac{m_{\Phi}^{2}}
{2}(F^{2}+G^{2})\nonumber\\
&-&\frac{\lambda}{\sqrt{2}}A\left[(\vec{X}^{2}-\vec{Y}^{2})F -
2G\vec{X}\cdot\vec{Y}\right].\nonumber
\eeqra
\begin{flushright}
(A.2)
\end{flushright}
Since $V_{N\Phi}^{tree}$ is now invariant under an $O(3)\otimes U(1)_{L}$
symmetry, the
possible vacua lay on $O(3)\otimes U(1)_{L}$ orbits. In particular, by
an appropriate $O(3)$ rotation, it is
not restrictive to consider the following vacuum

\[
F\neq 0,~~~G\neq 0, ~~~~~~~~~~~~~~
\left(\begin{array}{cc}
0&0\\
X_{2}& 0\\
X_{3}& Y_{3}
\end{array}\right).
\]
\begin{flushright}
(A.3)
\end{flushright}
The stationarity conditions read
\beqra
\partial_{X_{2}}V_{N\Phi}^{tree}&=& \left[2\lambda^{2}(F^{2}+G^{2})+\lambda^{2}
(\vec{X}^{2}-\vec{Y}^{2}) +m_{N}^{2}
-\sqrt{2}\lambda A F\right]X_{2}=0,\nonumber\\
\partial_{X_{3}}V_{N\Phi}^{tree}&=& \left[2\lambda^{2}(F^{2}+G^{2})+\lambda^{2}
(\vec{X}^{2}-\vec{Y}^{2}) +m_{N}^{2}
-\sqrt{2}\lambda A F\right]X_{3}+ 2\lambda^{2}(\vec{X}\cdot
\vec{Y})Y_{3}\nonumber\\
&+&\sqrt{2}\lambda A G Y_{3}=0,\nonumber\\
\partial_{Y_{3}}V_{N\Phi}^{tree}&=& \left[2\lambda^{2}(F^{2}+G^{2})-\lambda^{2}
(\vec{X}^{2}-\vec{Y}^{2}) +m_{N}^{2}
+\sqrt{2}\lambda A F\right]Y_{3}+ 2\lambda^{2}(\vec{X}\cdot
\vec{Y})X_{3}\nonumber\\
&+&\sqrt{2}\lambda A G Y_{3}=0,\nonumber\\
\partial_F V_{N\Phi}^{tree}&=& \left[2\lambda^{2}
(\vec{X}^{2}+\vec{Y}^{2}) +m_{\Phi}^{2}\right]F
-\frac{\lambda A}{\sqrt{2}}(\vec{X}^{2}-\vec{Y}^{2})=0\nonumber\\
\partial_{G}V_{N\Phi}^{tree}&=&\left[2\lambda^{2}
(\vec{X}^{2}+\vec{Y}^{2}) +m_{\Phi}^{2}\right]G+\sqrt{2}\lambda A (\vec{X}\cdot
\vec{Y})=0\nonumber.
\eeqra
\begin{flushright}
(A.4)
\end{flushright}
We discuss separately the two cases $X_{2}\neq 0$ and $X_{2}=0$.

$i)$ $X_{2}\neq 0$: we can set $G=0$ by a $U(1)_{L}$ rotation and
from the stationarity conditions we get
\[
(\vec{X}\cdot\vec{Y})Y_{3}=0.
\]
\begin{flushright}
(A.5)
\end{flushright}

If $Y_{3}\neq 0$, then $X_{3}=0$ and from the eqs. (A.4) we get
\[
F^{2}=\left(\frac{-m_{N}^{2}}{2\lambda^{2}}\right),
\]
\begin{flushright}
(A.6)
\end{flushright}
from which it is seen that $m_{N}^{2}$ must be negative. Similarly,
stationarity
conditions give
\[
\frac{A}{\lambda^{2}}\sqrt{-m_{N}^{2}}=\vec{X}^{2}-\vec{Y}^{2}\leq
\vec{X}^{2}+\vec{Y}^{2}=\frac{A^{2}-m_{\Phi}^{2}}{2\lambda^{2}}.
\]
\begin{flushright}
(A.7)
\end{flushright}
The above expressions implies the hierarchy
\[
m_{N}^{2}\leq m_{\Phi}^{2}< A^{2}.
\]
\begin{flushright}
(A.8)
\end{flushright}
Correspondingly, the value of $V_{N\Phi}^{tree}$ at the minimum is
\[
V_{N\Phi}^{tree}=
\frac{1}{2\lambda^{2}}m_{N}^{2}\frac{A^{2}-m_{\Phi}^{2}}{2}<0.
\]
\begin{flushright}
(A.9)
\end{flushright}
However, this minimum breaking $R$-parity is forbidden by RGE's, see eq. (22),
which give $m_{\Phi}^{2}<m_{N}^{2}$.

On the other hand, if $X_{2}\neq 0$ and $Y_{3}=0$, using the $O(3)$ symmetry,
we can take all the fields, except $F$ and $X_{3}$, to be vanishing. In this
case the original $O(3)\otimes U(1)_{L}$ breaks to $O(2)$, giving rise
to one Majoron and two familons. The minimization of the effective
potential is similar to that discussed in the text in sect. 2.

$ii)$ $X_{2}=0$: using a $U(1)_{L}$ rotation, one can choose $Y_{3}=0$
($F$, $G \neq 0$).
Stationarity conditions imply
\[
\sqrt{2}\lambda A G X_{3}=0,
\]
\begin{flushright}
(A.10)
\end{flushright}
which gives $G=0$ for $X_{3}\neq 0$ (the opposite case $X_{3}=0$ and $G\neq 0$
does not give a minimum). The analysis of such a case reduces to the previously
analyzed case $i)$ with $Y_{3}=0$.

\vskip 0.4 cm
\begin{description}
\centerline{\Large{\bf Figure Captions}}
\vskip 0.3 cm
\item[1)] Allowed region in the $({\rm Log}_{10} \lambda,~ A)$ parameter space
for the breaking of $R$-parity (denoted by dots), in the case
$\lambda_{1}=\lambda_{2}=0$, $\lambda_{3}=\lambda$. The region above the
continuous line is forbidden by the $|A|<5\tilde{m}$ condition imposed at the
GUT scale. The regions below the dotted and the dashed lines correspond to
positive values for $m_{\Phi}^{2}$ and $m_{N}^{2}$, respectively. $A$ is given
in units of $\tilde{m}$.
\item[2)] Same as in fig. $1)$, but for $\lambda_{1}=\lambda_{2}=
\lambda_{3}=\lambda$.
\end{description}

\vspace{1.5 truecm}
\baselineskip=20pt

\begin{thebibliography}{20}
\bibitem{seesaw}M. Gell-Mann, P. Ramond, and R. Slansky, in {\it
Supergravity}, F. van Nieuwenhuizen and D. Freedman, eds.
(North Holland, Amsterdam 1979) pag. 315; T. Yanagida, in {\it
Proceedings of the Workshop on Unified Theory and Baryon Number
of the Universe}, KEK, Japan, 1979.
\bibitem{maj}G.B. Gelmini and M. Roncadelli, Phys. Lett. {\bf 99B}
(1981) 411.
\bibitem{majsing}Y. Chikashige, R.N. Mohapatra, and R. Peccei,
Phys. Lett. {\bf 98B} (1981) 265.
\bibitem{rbreak}L.J. Hall and M. Suzuki, Nucl. Phys. {\bf B231} (1984)
419; S. Dawson, Nucl. Phys. {\bf B261} (1985) 297; S. Dimopoulos
and L.J. Hall, Phys. Lett. {\bf B207} (1987) 210; V. Barger, G.F.
Giudice, and T. Han, Phys. Rev. {\bf D40} (1989) 2987; S. Dimopoulos,
R. Esmailzadeh, L.J. Hall, J.P. Merlo and G.S. Starkmen Phys. Rev. {\bf D41}
(1990) 2099, K. Enqvist, A. Masiero and A. Riotto, Nucl. Phys. {\bf B373}
(1992) 95.
\bibitem{rspont}C. Aulak and R. Mohapatra, Phys. Lett. {\bf 119B}
(1983) 136; J. Ellis, {\it et al.}, Phys. Lett. {\bf B150} (1985)
142; G.G. Ross and J.W.F. Valle, Phys. Lett. {\bf B151} (1985) 375.
\bibitem{lep} F. Dydak, {\it Proc. XXV HEP conference}, Singapore, 1990.
\bibitem{valle}A. Masiero and J.W.F. Valle, Phys. Lett. {\bf B251}
(1990) 273; J.C. Rom\~ao, C.A. Santos, and J.W.F. Valle, preprint
FTUV/91-60 (1991).
\bibitem{sphal} V.A. Kuzmin, V.A. Rubakov and M.E. Shaposhnikov, Phys.
Lett. {\bf B155} (1985) 36.
\bibitem{campbell} B. Campbell, S. Davidson, J. Ellis and K.A. Olive,
Phys. Lett. {\bf B256} (1991) 457 and CERN/TH 6208/91 preprint;
W. Fishler, G.F. Giudice, R.G. Leigh and S. Paban, Phys. Lett. {\bf B258}
(1991) 45.
\bibitem{hall} S. Dimopoulos and L.J. Hall, Phys. Lett. {\bf B196} (1987) 135;
J. Cline and S. Raby, Phys. Rev. {\bf D43} (1991) 1781.
\bibitem{riotto} A. Masiero and A. Riotto, DFPD 92/TH/22 and SISSA-78/92/AP,
to appear in Phys. Lett. {\bf B}.
\bibitem{dreiner} H. Dreiner and G.G. Ross, Oxford preprint OUTP 92-08P.
\bibitem{color}J.M. Fr\`ere, D.R.T. Jones, and S. Raby, Nucl. Phys.
{\bf B222} (1983) 11; M. Claudson, L.J. Hall, and I. Hinchliffe,
Nucl. Phys. {\bf B228} (1983) 501; J.-P. Derendinger and C.A. Savoy,
Nucl. Phys. {\bf B237} (1984) 307; M. Drees, M. Gl\"uck, and K.
Grassie, Phys. Lett. {\bf B157} (1985) 164; J.F. Gunion, H.E. Haber,
and M. Sher, Nucl. Phys. {\bf B306} (1988) 1; H. Komatsu, Phys.
Lett. {\bf B215} (1988) 323.
\bibitem{raff}G.G. Raffelt, Phys. Rep. {\bf 198} (1990) 1.
\bibitem{valtau}J.C. Rom\~ao, N. Rius, and J.W.F. Valle, Nucl.
Phys. {\bf B363} (1991) 369.
\bibitem{dolan} L. Dolan and R. Jackiw, Phys. Rev. {\bf D9} (1974) 3320,
S. Weinberg, Phys. Rev. {\bf D9} (1974) 3357.
\bibitem{fukugita} Y. Kondo, I. Umemura and K. Yamamoto, Phys. Lett. {\bf B263}
(1991) 93.
\bibitem{ibanez} L.E. Ib\'{a}\~{n}ez and G.G. Ross, Nucl. Phys. {\bf B368}
(1992) 3.
\end{thebibliography}
\end{document}